\documentclass[prl,twocolumn,showpacs,amsmath]{revtex4}
\usepackage{bm}
\newcommand{\eps}{\varepsilon}
\newcommand{\rmd}{\mathrm d}
\newcommand{\rmi}{\mathrm i}
\newcommand{\mat}[1]{\tensor{#1}}
\newcommand{\bracket}[1]{\langle #1 \rangle}
\newcommand{\Bra}[1]{\Bigl \langle #1 \Bigr \rvert}
\newcommand{\Ket}[1]{\Bigl \lvert #1 \Bigr \rangle}
\newcommand{\ket}[1]{\lvert #1 \rangle}
\newcommand{\dpar}[2]{\frac{\partial #1}{\partial #2}}

\begin{document}

\title{Berry phase correction to electron density of states in solids}
\author{Di Xiao}
\affiliation{%
Department of Physics,
The University of Texas,
Austin, TX 78712-0264, USA}
\author{Junren Shi}
\affiliation{%
Department of Physics,
The University of Texas,
Austin, TX 78712-0264, USA}
\author{Qian Niu}
\affiliation{%
Department of Physics,
The University of Texas,
Austin, TX 78712-0264, USA}

\date{February 14, 2005}

\begin{abstract}
Liouville's theorem on the conservation of phase space volume is
violated by Berry phase in the semiclassical dynamics of Bloch
electrons.  This leads to a modification of the phase space density of
states, whose significance is discussed in a number of examples: field
modification of the Fermi-sea volume, connection to the anomalous Hall
effect, and a general formula for orbital magnetization.  The
effective quantum mechanics of Bloch electrons is also sketched, where
the modified density of states plays an essential role.
\end{abstract}


\maketitle

Semiclassical dynamics of Bloch electrons in external fields has
provided a powerful theoretical framework to account for various
properties of metals, semiconductors and insulators
\cite{ashcroft1976}.  In recent years, it has become increasingly
clear that essential modification of the semiclassical dynamics is
necessary for a proper understanding of a number of phenomena.  It was
known earlier that global geometric phase effects
\cite{berry1984,gphase2003} on Bloch states are very important for
insulators in our understanding of the quantum Hall effect
\cite{thouless1982}, quantized adiabatic pumps \cite{thouless1983},
and electric polarization \cite{kingsmith1993,resta1994}.  It was
shown \cite{chang1996,sundaram1999} later that geometric phase also
modifies the local dynamics of Bloch electrons and thus affects the
transport properties of metals and semiconductors.  Recently these
ideas have been successfully applied to the anomalous Hall effect in
ferromagnetic semiconductors and metals \cite{jungwirth2002, fang2003,
yao2004,haldane2004}, as well as spin transport
\cite{murakami2003,sinova2004}.

In this Letter, we reveal a general property of the Berry phase
modified semiclassical dynamics which has been overlooked so far: the
violation of Liouville's theorem for the conservation of phase space
volume.  Liouville's theorem was originally established for standard
classical Hamiltonian dynamics, and its importance cannot be over
emphasized as it serves as a foundation for classical statistical
physics.  The Berry phase makes, in general, the equations of motion
non-canonical~\cite{kuratsuji1988, littlejohn1991, chang1996,
sundaram1999, panati2003}, rendering the violation of
Liouville's theorem.  Nevertheless, we are able to remedy the
situation by modifying the density of states in the phase space.

This modified phase-space density of states enters naturally in the
semiclassical expression for the expectation value of physical
quantities, and has profound effects on equilibrium as well as
transport properties.  We demonstrate this with several examples.
First, we consider a Fermi sea of electrons in a weak magnetic field,
and show that the Fermi sea volume can be changed linearly by the
field.  Second, we show how the Berry phase formula for the intrinsic
anomalous Hall conductivity may be derived from equilibrium
thermodynamics using the St{\v r}eda formula \cite{streda1982}.
Third, we provide a general derivation of an orbital-magnetization
formula which is convenient for first-principles calculations.

In addition, we present an effective quantum mechanics for Bloch
electrons in solids by quantizing the semiclassical dynamics with the
geometric phase.  The density of states enters in a nontrivial manner
into the commutators of the phase space coordinates, and relates
directly to the minimal uncertainty volume in the phase space.

To begin with, we write down the semiclassical equations of motion for
a Bloch electron in weak electric and magnetic fields
\cite{sundaram1999}
\begin{subequations}
\label{EOM}
\begin{align}
\dot{\bm r} &= \frac{1}{\hbar} \dpar{\eps_n(\bm k)}{\bm k}
- \dot{\bm k} \times \bm \Omega_n(\bm k) \;, \\
\hbar \dot{\bm k} &= -e \bm E(\bm r) - e\dot{\bm r} 
\times \bm B(\bm r) \;,
\end{align}
\end{subequations}
where $\bm \Omega_n(\bm k)$ is the Berry curvature of electronic Bloch
states defined by $\bm \Omega_n(\bm k) = \rmi \bracket{\bm \nabla_{\bm
k} u_n(\bm k)|\times|\bm \nabla_{\bm k} u_n(\bm k)}$ with
$\ket{u_n(\bm k)}$ being the periodic part of Bloch waves in the $n$th
band, $\eps_n(\bm k)$ is the band energy with a correction due to the
orbital magnetic moment [see Eq.~\eqref{energy} and above].  For
crystals with broken time-reversal symmetry (such as ferromagnetic
materials) or spatial inversion symmetry (such as GaAs), the Berry
curvature $\bm\Omega_n(\bm k)$ is nonzero.

To show the violation of Liouville's theorem, we consider the
time evolution of a volume element $\Delta V = \Delta \bm r\Delta \bm
k$ in the phase space.  The equation of motion for $\Delta V$ is given
by $(1/\Delta V) \rmd \Delta V / \rmd t = \bm \nabla_{\bm r} \cdot
\dot{\bm r} + \bm \nabla_{\bm k} \cdot \dot{\bm k}$~\cite{reichl1980}.
A straightforward but somewhat tedious calculation shows that the
right hand side is equal to $-\rmd \ln (1 + e\bm B \cdot \bm
\Omega/\hbar)/\rmd t$, which is a total time derivative.  Therefore we
can solve for the time evolution of the volume element and obtain
\begin{equation}
\Delta V = \Delta V_0/(1 + e \bm B \cdot \bm \Omega_n / \hbar) \;.
\label{volume}
\end{equation}
The fact that the Berry curvature is generally $\bm k$ dependent (and
the magnetic field can also depend on $\bm r$) implies that the phase
space volume element changes during time evolution of the state
variables $(\bm r, \bm k)$.

Nevertheless, we have a remedy to this breakdown of Liouville's
theorem.  Equation~\eqref{volume} shows that the volume element is a
local function of the state variables (through the magnetic field and
the Berry curvature) and has nothing to do with the history of time
evolution.  We can thus introduce a modified density of states
\begin{equation}
D_n(\bm r, \bm k) = (2\pi)^{-d} 
(1 + e \bm B \cdot \bm \Omega_n / \hbar) \;,
\label{DOS}
\end{equation}
such that the number of states in the volume element, $D_n(\bm r, \bm
k) \Delta V$, remains constant in time, where $d$ is the spatial
dimensionality of the system.  The prefactor $(2\pi)^{-d}$ is obtained
by demanding that the density of states $D_n(\bm r, \bm k)$ reduces to
the conventional form when the Berry curvature vanishes.  As will be
shown later, this density of states corresponds to the minimal quantum
uncertainty volume of the state variables.  Therefore, it does serve
as the semiclassical measure for the number of quantum states per unit
volume in the phase space. Based on this understanding, we write the
classical phase-space probability density as
\begin{equation}
\rho_n(\bm r, \bm k, t) = D_n(\bm r, \bm k) f_n(\bm r, \bm k, t)\; ,
\label{rho}
\end{equation}
with $f_n(\bm r, \bm k, t)$ being the occupation number of the state
labeled by $(\bm r, \bm k)$.  Probability conservation demands that
$\rho_n(\bm r, \bm k, t)$ satisfies the continuity equation in phase
space.  On the other hand, our density of states satisfies $\rmd D_n /
\rmd t = -(\bm \nabla_{\bm r} \cdot \dot{\bm r} + \bm \nabla_{\bm k}
\cdot \dot{\bm k})D_n$.  It then follows that the occupation number
introduced above has the desired property of being invariant along the
trajectory, i.e., $\rmd f_n /\rmd t=0$~\cite{boltzmann}.

We can thus write the real space density of a physical observable
$\hat O$ in the form~\cite{culcer2004}
\begin{equation}
\bar{O}(\bm R) = \sum_n \int \rmd\bm k D_n(\bm r, \bm k) 
f_n(\bm r, \bm k, t) 
\langle \hat{O}\delta(\hat{\bm r}-\bm R) \rangle_{\bm r \bm k n}
\label{average_general}
\end{equation}
where $\langle \cdots \rangle_{\bm r \bm k n}$ denotes the expectation
value in the wave-packet state centered at ($\bm r$, $\bm k$) with the
band index $n$.  In the spatially homogeneous case, it reduces to:
\begin{equation}
\bar{O} = \sum_n \int \rmd\bm k D_n(\bm k) 
f_n(\bm k) O_n(\bm k)  \;,
\label{average}
\end{equation}
where $O_n(\bm k)$ is the expectation value of $\hat O$ in a Bloch
state.  For simpler notation, we will drop the band index $n$ and
assume that the integral over $\bm k$ includes the sum over $n$.

We now discuss the magnitude of the correction term $e\bm B \cdot \bm
\Omega/\hbar$ to the density of states in Eq.~\eqref{DOS}.  The Berry
curvature for several materials has been calculated before using
first-principles method~\cite{fang2003,yao2004}.  Over large regions
of the Brillouin zone, its magnitude is on the order of $a^2$ with $a$
being the lattice constant.  Thus, $e\bm B\cdot\bm\Omega/\hbar\sim
eBa^2/\hbar$ is the ratio of the magnetic flux through a unit cell to
the magnetic flux quantum, and can be $10^{-2}$ to $10^{-3}$ for a
magnetic field of 1 tesla.  In the vicinity of some isolated points,
the Berry curvature can be several orders of magnitude higher, leading
to bigger effects for measurement.  In the following, we will present
a number of applications of our formula Eq(6).

In our first example, we consider the quantity of electron density and
show that the Fermi sea volume can be changed linearly by a magnetic
field when the Berry curvature is non-zero.  Assuming zero temperature
and using Eq.~\eqref{DOS}, we have the electron density as
\begin{equation}
n_e = \int^\mu
\frac{\rmd\bm k}{(2\pi)^d} 
\left(1+\frac{e\bm B \cdot \bm\Omega}{\hbar} \right) \;,
\label{density}
\end{equation}
where the upper limit means that the integral is over states with
energies below the chemical potential $\mu$.  Noting that the electron
density is fixed by the background charge density, we conclude that
the Fermi volume  must change with the magnetic field.  To first
order, this change is given by
\begin{equation}
\delta V_F = - \int^{\mu_0}
\rmd\bm k\,\frac{e\bm B\cdot \bm \Omega}{\hbar} \;.
\label{FermiV}
\end{equation}
We note that while Landau levels make the Fermi sea volume oscillate
with the field, the effect described above gives an overall shift on
average.  Such a shift has important implications to those
Fermi-surface related behaviors such as transport properties.  For
instance, in metals, it can induce a magnetoresistance linearly
depending on the magnetic field.  On the other hand, in band
insulators, the $\bm k$ space is limited to the Brillouin zone.
Electrons must populate a higher band if $(e/\hbar)\int_\text{BZ}
\rmd\bm k\, \bm B\cdot \bm \Omega$ is negative.  When this quantity is
positive, holes must appear at the top of the valence bands.
Discontinuous behavior of physical properties in a magnetic field is
therefore expected for band insulators with non-zero integral of the
Berry curvatures (Chern numbers).

In our second example, we show a connection between our phase space
density of states to the intrinsic anomalous Hall effect, which is due
to spin-orbit coupling in the band structure of a ferromagnetic
crystal.  In the context of the quantum Hall effect, St{\v r}eda
derived a formula relating the Hall conductivity to the field
derivative of the electron density at a fixed chemical potential
\cite{streda1982}, $\sigma_{xy}=-e(\partial n_e/\partial B_z)_\mu$.
There is a simple justification of this relation by a thermodynamic
argument by considering the following adiabatic process in two
dimensions. A time dependent magnetic flux generates an electric field
with an emf around the boundary of some region; and the Hall current
leads to a net flow of electrons across the boundary and thus a change
of electron density inside.  This argument can be straightforwardly
applied to the case of anomalous Hall effect and to three dimensions.
By taking the derivative of the electron density~\eqref{density} with
respect to $\bm B = B\hat{\bm z}$ at fixed chemical potential, we find
that
\begin{equation}
\sigma_{xy} = -\frac{e^2}{\hbar}
\int^\mu \frac{\rmd\bm k}{(2\pi)^d} \Omega_z \;.
\label{sigma}
\end{equation}
This is an intrinsic effect because it is independent of scattering,
and thus differs from conventional skew scattering and side jump
mechanisms~\cite{jungwirth2002,fang2003,yao2004,haldane2004}.
 
As a third example of application, we now derive a semiclassical
formula for orbital magnetization.  In the semiclassical picture, a
Bloch electron is modeled by a wave packet in a Bloch band, which is
found to rotate about its center of mass in general, yielding an
intrinsic magnetic moment given by $\bm m(\bm k) = -\rmi (e/2\hbar)
\bracket{\bm \nabla_{\bm k}u| \times [\hat{H}_0(\bm k) - \eps_0(\bm
k)]|\bm \nabla_{\bm k}u}$, where $\hat{H}_0$ is the
Hamiltonian~\cite{m-moment}.  In the presence of a weak magnetic field
$\bm B$, the electron band structure energy $\eps_0(\bm k)$ (which may
already include Zeeman energy from spin magnetization) acquires a
correction term from this intrinsic orbital moment
\cite{chang1996,sundaram1999}, $\eps(\bm k) = \eps_0(\bm k) - \bm
m(\bm k) \cdot \bm B$.  For an equilibrium ensemble of electrons, the
total orbital magnetization can be found from the total energy, which
is given by Eq.~\eqref{average} as,
\begin{equation}
E = \int^\mu \frac{\rmd\bm k}{(2\pi)^d} 
\Bigl(1 + \frac{e \bm B \cdot \bm \Omega}{\hbar} \Bigr)
\Bigl(\eps_0(\bm k) - \bm m(\bm k) \cdot \bm B \Bigr) \;.
\label{energy}
\end{equation}
Taking differential of $E$ with respect to $\bm B$, we obtain the
magnetization at zero magnetic field to be
\begin{align}
\bm M &= \int^{\mu_0} \frac{\rmd\bm k}{(2\pi)^d} \Bigl(\bm m(\bm k) 
+ \frac{e\bm \Omega}{\hbar}\bigr[\mu_0 - \eps_0(\bm k)\bigr] \Bigr) \\
&= \frac{e}{2\hbar} \int^{\mu_0} \frac{\rmd \bm k}{(2\pi)^d} \rmi
\Bra{\dpar{u}{\bm k}} \times [2\mu_0 - \eps_0(\bm k) - \hat{H}_0]
\Ket{\dpar{u}{\bm k}} \;. \nonumber
\label{magnetization}
\end{align}
In the upper line of the above expression, the first term is the
contribution from the intrinsic orbital moment of each Bloch electron,
and the second term comes from the explicit field dependence of the
density of states and the resulting change in the Fermi volume in
Eq.~\eqref{FermiV}.  We expect this effect to be important in
ferromagnetic materials with strong spin-orbit coupling.

Gat and Avron obtained an equivalent result for the special case of
Hofstadter model~\cite{gat2003}.  Our derivation provides a more
general formula that is applicable to other systems.  Following the
discussions on band insulators in our first example, there will be a
discontinuity of the orbital magnetization if the integral of the
Berry curvature over the Brillouin zone, or the anomalous Hall
conductivity, is non-zero and quantized. Depending on the direction of
the field, the chemical potential $\mu_0$ in the above formula should
be taken at the top of the valence bands or the bottom of the
conduction bands.  The size of the discontinuity is given by the
quantized anomalous Hall conductivity times $E_g/e$, where $E_g$ is
the energy gap.  For insulators with zero Chern numbers, the orbital
magnetization can be directly evaluated from Wannier functions, with
result consistent with our general formula \cite{resta2005}. Our
general formula can also be derived from a full quantum mechanical
linear response analysis \cite{shi2005}.

The central result of this paper, equation~\eqref{DOS}, can be
extended to the more general case when Berry curvature includes the
components of $\mat{\Omega}^{\bm k \bm r}$ as well as $\mat{
\Omega}^{\bm k\bm k}$ and $\mat{\Omega}^{\bm r \bm
r}$~\cite{sundaram1999}.  In this case, we introduce the Berry
curvature in phase space,
\begin{equation}
\mat{\bm \Omega} = \begin{pmatrix} \mat{\Omega}^{\bm r\bm r} &
\mat{\Omega}^{\bm r\bm k} \\ \mat{\Omega}^{\bm k\bm r} & 
\mat{\Omega}^{\bm k\bm k} \end{pmatrix} \;,
\end{equation}
where each block is a $3\times3$ matrix; $\mat{\Omega}^{\bm r\bm
k} = -(\mat{\Omega}^{\bm k\bm r})^T$.  The phase space density
of states then reads,
\begin{equation}
D = (2\pi)^{-d} \sqrt{\det(\mat{\bm\Omega} -  \mat{\mathbf J})} \;.
\label{DOS-gen}
\end{equation}
with $\mat{\mathbf J} = \bigl(\begin{smallmatrix} 0 & \mat{\mathrm I}
\\- \mat{\mathrm I} & 0\end{smallmatrix}\bigr)$.  In the special case
of electromagnetic perturbations with $\mat{\Omega}^{\bm k\bm
k}_{ab}=\epsilon_{abc}\Omega_c$, $\mat{\Omega}^{\bm r\bm r}_{ab}=- (e/
\hbar) \epsilon_{abc} B_c$ and $\mat{\Omega}^{\bm k \bm r}=0$, it
reduces to \eqref{DOS}.  On the other hand, when either
$\mat{\Omega}^{\bm k \bm k}$ or $\mat{\Omega}^{\bm r \bm r}$ vanishes,
it has a simpler form
\begin{equation}
D = (2\pi)^{-d}  \det(\mat{\mathrm I}
 - \mat{\Omega}^{\bm r\bm k})  \;.
\end{equation}
This result has found application in the study of spin-force induced
charge-Hall effect \cite{zhang2004}.

Finally, we show how the density of states emerges naturally in the
effective quantum mechanics of Bloch electrons.  Although our system
is not canonical, it can nevertheless be quantized following a
standard procedure developed for non-holonomic systems with second
class constraints~\cite{senjanovic1976,gitman1990}.  First, one
redefines the Poisson bracket $\{f, g\}^* = (\partial f / \partial
\xi^a) M_{ab} (\partial g / \partial \xi^b)$, where $\xi^a$ are the
components of phase space coordinates $\bm \xi \equiv (\bm r,\; \bm
k)$ and $\mat{\mathbf M} = (\mat{\bm\Omega} - \mat{\mathbf J})^{-1}$.
Our equations of motion~\eqref{EOM} can then be written as
$\dot{\xi}^a = \{\xi^a, \eps\}^*$, where the energy $\eps(\bm \xi)$
plays the role as the Hamiltonian function.  Then, one promotes the
Poisson brackets into quantum commutators:
\begin{equation}
[\hat{\xi}^a, \hat{\xi}^b] = \rmi M_{ab} \;,
\label{commutator_gen}
\end{equation}
where $\hat{\xi}^a$ is the quantum operator corresponding to the phase
space coordinates.  It then follows that a phase space point
acquires a minimal uncertainty volume given by \cite{robertson1934}
\begin{equation}
\min \Bigl(\prod_a \Delta \xi^a\Bigr) =
2^{-d} \Bigl[\det(\mat{\bm \Omega} - \mat{\mathbf J})\Bigr]^{-1/2} \;.
\end{equation}
This can be understood as the phase space volume occupied by a single
quantum state, therefore Eq.~\eqref{DOS-gen}, which is proportional to
the reciprocal of this volume, can naturally be regarded as the
semiclassical expression for the number of quantum states per unit
volume in the phase space.

Equation~\eqref{commutator_gen} presents the effective quantum
mechanics of Bloch electrons.  As a demonstration for the validity of
the quantization scheme as well as the quantum effect of the phase
space density of states, we consider a simple toy model of two
dimensional electron system with a constant Berry curvature, subjected
to a uniform magnetic field.  The commutators read,
\begin{align}
&[\hat{x}, \hat{y}] = \rmi \frac{\Omega}{1 + (e/\hbar)B\Omega}
\;, \quad [\hat{k}_x, \hat{k}_y] = -\rmi \frac{(e/\hbar)B}{1 +
(e/\hbar)B\Omega} \;, \nonumber \\ 
&[\hat{x}, \hat{k}_x] = [\hat{y}, \hat{k}_y] =
\rmi \frac{1}{1 + (e/\hbar)B\Omega} \;.
\label{commutator}
\end{align}
In the absence of the Berry curvature, we reduce the problem to a
known case with the familiar nontrivial commutator $[\hat{k}_x,
\hat{k}_y] = -\rmi (e/\hbar) B$.  In the absence of the $ B$ field,
we have the nontrivial commutator $[\hat{x}, \hat{y}] = \rmi \Omega$
discussed extensively in the literature on non-commutative geometry.
It is interesting to see that in the presence of both fields, we do
not just have a combination of these nontrivial commutators.
Instead, we have a nontrivial density of states which enters into all
of the commutators.

Assuming $\eps(\bm k)=\hbar^2 {\bm k}^2/2m$, the system can be solved
algebraically to yield the energy spectrum and degeneracy.  We found
that the spectrum consists of a set of Landau levels with the
renormalized cyclotron frequency $ \omega_c = \omega_c^0/[1 +
(e/\hbar)B\Omega]$, where $\omega_c^0 = eB/m$ is the usual cyclotron
frequency~\cite{horvathy2005}.  At the same time, it is more important
to note that each Landau level still has the same degeneracy of $eB/h$
as in the absence of the Berry curvature.  It is known that this
degeneracy is directly related to the quantized Hall conductance
$e^2/h$ for a filled Landau level~\cite{hall}.  Had the density of
states not entered in the commutators, the Landau level degeneracy
would be modified, violating the topological requirement that the Hall
conductance for a filled Landau level is quantized.

Before closing, we note that the phase space density of states also
enters naturally in the alternative quantization scheme with Feynman
path integral. The $S$ matrix is calculated by~\cite{senjanovic1976}
\begin{equation}
\bracket{\text{out}|S|\text{in}} = \int \prod_t [D(\bm \xi) 
\rmd \bm \xi]
 \exp\Bigl[{\frac{\mathrm i}{\hbar} \int L \rmd t}\Bigr] \;.
\end{equation}
where $L$ is the Lagrangian for our system~\cite{sundaram1999},
\begin{equation}
L = \frac{1}{2}\dot{\xi}^a J_{ab} \xi^b - \eps(\bm \xi)
+ \dot{\xi}^a \mathcal A_a(\bm \xi) \;
\label{lagrangian}
\end{equation}
with $\mathcal A_a(\bm \xi) \equiv \rmi \bracket{u(\bm
\xi)|\nabla_\alpha u(\bm \xi)}$ being the phase space gauge potentials
associated with the Berry curvature field $\mat{\bm\Omega}$.

In summary, we have found a Berry phase correction to the phase space
density of states for Bloch electrons.  This correction emerges
naturally in both semiclassical and quantum mechanics of Bloch
electrons, and has profound effects on the equilibrium and transport
properties. Because of the fundamental change introduced by this
correction, it could have important implications on other aspects of
condensed matter physics, such as the Fermi liquid theory.  For
instance, in the presence of a magnetic field, interaction between
electrons can change the Fermi sea volume by modifying the Berry
curvature and thus the phase space density of states.

We acknowledge useful discussions with M.~C.~Chang, R.~Resta,
G.~Y.~Guo, Y.~G.~Yao, and D.~Vanderbilt.  This work is supported by
DOE grant DE-FG03-02ER45958.

\bibliography{journal,book,eprint,footnote}

\begin{thebibliography}{32}
\bibitem[{hal()}]{hall}
\bibinfo{note}{The Hall conductance is given by $\sigma = ne/B$, where $n$ is
  the electron density. The quantized value is obtained if we substitute the
  degeneracy density for $n$.}
\end{thebibliography}

\end{document}